# A Systematic Review of Machine Learning Methods for Multimodal EEG Data in Clinical Application

**Siqi Zhao**[1*], **Wangyang Li**[1*], **Xiru Wang**[1*], **Stevie Foglia**[2*], **Hongzhao Tan**[1*], **Bohan Zhang**[1], **Ameer Hamoodi**[2], **Aimee Nelson**[2,3], **Zhen Gao**[1,2]

[1] WBooth School of Engineering Practice and Technology, McMaster University, Hamilton, Ontario Canada
[2] School of Biomedical Engineering, McMaster University, Hamilton, Ontario, Canada
[3] Department of Kinesiology, McMaster University, Hamilton, Ontario, Canada

**Emails**:
foglias@mcmaster.ca, zhaos98@mcmaster.ca, nelsonaj@mcmaster.ca, gaozhen@mcmaster.ca

*These authors contributed equally to this work

**Abstract**

Machine learning (ML) and deep learning (DL) techniques have been widely applied to analyze electroencephalography (EEG) signals for disease diagnosis and brain-computer interfaces (BCI). The integration of multimodal data has been shown to enhance the accuracy of ML and DL models. Combining EEG with other modalities can improve clinical decision-making by addressing complex tasks in clinical populations. This systematic literature review explores the use of multimodal EEG data in ML and DL models for clinical applications. A comprehensive search was conducted across PubMed, Web of Science, and Google Scholar, yielding 16 relevant studies after three rounds of filtering. These studies demonstrate the application of multimodal EEG data in addressing clinical challenges, including neuropsychiatric disorders, neurological conditions (e.g., seizure detection), neurodevelopmental disorders (e.g., autism spectrum disorder), and sleep stage classification. Data fusion occurred at three levels: signal, feature, and decision levels. The most commonly used ML models were support vector machines (SVM) and decision trees. Notably, 11 out of the 16 studies reported improvements in model accuracy with multimodal EEG data. This review highlights the potential of multimodal EEG-based ML models in enhancing clinical diagnostics and problem-solving.

Keywords: EEG, Deep Learning, multimodal, clinical population

## Introduction

### 1.1 EEG background

Electroencephalography (EEG) is a non-invasive method used to capture and record the brain's electrical activity through electrodes placed on the scalp. These electrodes measure voltage fluctuations caused by ionic currents within neurons, enabling the real-time recording of brain wave patterns (1) The process involves amplifying these electrical signals and displaying them as waveforms, which can be analysed to understand different states of brain activity, such as wakefulness, sleep, and various levels of consciousness. EEG signals are divided into different frequency bands, including delta, theta, alpha, beta, and gamma waves, each associated with specific cognitive and physiological states. This capability to monitor and interpret brain waves in real time has made EEG a foundational tool in neurology and cognitive neuroscience.

EEG's applications go well beyond merely observing brain activity; it is essential for diagnosing and treating numerous neurological and psychiatric conditions. For example, EEG is crucial in identifying abnormal brain wave patterns associated with epilepsy (2), assisting in both diagnosis and monitoring of the disorder. It can locate epileptic activity, which is vital for surgical planning in drug-resistant epilepsy cases. EEG is also invaluable in diagnosing sleep disorders such as insomnia, narcolepsy, and sleep apnea by providing insights into sleep architecture and disturbances (3). The versatility and diagnostic power of EEG have significantly advanced our ability to understand and treat a wide range of brain-related disorders, thereby improving patient outcomes and quality of life. Importantly, in recent years, EEG as a modailty has been enhanced by combing it use with other imaging and data modalities to improve the sensitivity and specificity of data.

*1.2 Machine Learning Techniques for EEG Signals*

Building on the foundational understanding of EEG data and its significance, it is essential to explore the advanced methodologies that have been developed to analyse and address the challenges associated with EEG signals. Machine Learning (ML) and Deep Learning (DL) methods have become increasingly prominent in analyzing and resolving EEG data and related problems due to their ability to automatically extract complex patterns from large datasets. ML is a broad field of study that focuses on developing algorithms that can learn from and make predictions on data. DL has revolutionized the field of ML by enabling the automatic extraction of intricate structures in large datasets (4). DL methods significantly outperform traditional approaches in various domains, such as speech recognition, image recognition, and natural language processing, by utilizing multiple layers of representation to learn from raw data (5). Various EEG problems that ML or DL methods aim to address include epileptic seizure detection, sleep stage classification, mental state and cognitive workload assessment, emotion recognition, and brain-computer interface (BCI) applications (6,7). Several DL techniques have been employed to tackle these challenges. Convolutional Neural Networks (CNNs) are widely used for their ability to capture spatial features in EEG signals, making them effective in tasks like seizure detection and sleep stage classification (8,9). Recurrent Neural Networks (RNNs), particularly Long Short-Term Memory (LSTM) (10) networks are utilized to analyse the temporal dependencies in EEG data, which is crucial for understanding time-series data and predicting mental states (11). Autoencoders and Variational Autoencoders (VAEs) have been applied for feature extraction and noise reduction, enhancing the quality of EEG signal analysis (12,13). Furthermore, hybrid models combining CNNs and RNNs are increasingly popular for leveraging both spatial and temporal information, providing robust solutions for complex EEG-based applications. These DL methods collectively contribute to significant advancements in the interpretation and utilization of EEG data in both clinical and non-clinical settings.

*1.3 EEG Multimodal data*

Multimodal data integration involves the simultaneous collection and analysis of data from multiple sources, providing a more holistic view of brain function and its interactions with physiological and behavioral states. This approach leverages the strengths of different neurophysiological tools to overcome the limitations of individual modalities, resulting in richer and more comprehensive datasets (14).

Neurophysiological data, including EEG, functional magnetic resonance imaging (fMRI), eye tracking, skin conductance, and heart rate monitoring, have enabled significant advancements in both academic and commercial neuromarketing research (15). EEG is frequently integrated with other modalities, such as fMRI, eye tracking, and various bio-signals (e.g., heart rate and skin conductance), to achieve a more comprehensive understanding of brain function and its interplay with physiological and behavioral states.

In recent studies, several modalities are commonly combined with EEG to enrich data analysis and interpretation, including functional near-infrared spectroscopy (fNIRS), eye tracking, electromyography (EMG), electrocardiography (ECG), electrooculography (EOG), and magnetic resonance imaging (MRI), providing a multidimensional perspective on brain function and its correlation with physiological and behavioral responses (16–20).

*1.4 Multimodal EEG Data & Disease*

Multimodal EEG data refers to the integration of EEG with other physiological and neuroimaging modalities such as fNIRS, MRI, positron emission tomography (PET), and various biosignals, including ECG and respiratory signals. The primary advantage of using multimodal data lies in its enhanced ability to capture a more comprehensive picture of brain activity, thus enabling more accurate diagnosis and tailored treatment approaches for neurological disorders.

The integration of EEG with various physiological and neuroimaging modalities has been extensively explored in the study of multiple neurological and psychiatric disorders, including epilepsy, autism spectrum disorder (ASD), depression, anxiety disorders, and cognitive impairments. Each modality adds unique value, providing structural, functional, and behavioral insights that lead to more accurate assessments and understanding of these conditions.

Multimodal EEG data has proven highly effective in advancing the detection and management of various neurological and psychiatric conditions. For epilepsy, combining EEG with fNIRS enhances seizure detection by capturing both electrical and hemodynamic responses during seizure events, thereby increasing diagnostic accuracy in clinical monitoring settings (21,22). Integrating EEG with EMG and accelerometer data further improves the detection of tonic-clonic seizures, enabling robust, long-term monitoring for high-risk patients (23). These multimodal systems offer a comprehensive approach to seizure detection that surpasses traditional, unimodal methods in reliability and effectiveness.

In ASD detection, multimodal EEG data has shown promise by integrating EEG with eye-tracking and behavioral metrics such as eye fixation and facial expression. This approach captures neural and behavioral markers, enabling a more nuanced assessment of atypical social responses that are characteristic of ASD. As a result, early ASD detection accuracy has significantly improved, supporting timely intervention (17,24). Similarly, multimodal EEG data combined with heart rate variability (HRV) helps identify arousal states in anxiety disorders, especially in virtual reality exposure therapy, where these signals guide interventions to reduce stress and enhance patient outcomes (18).

For cognitive impairments and sleep disorders, multimodal EEG data also plays a critical role. In mild cognitive impairment (MCI) and Alzheimer's disease, EEG combined with structural MRI offers an improved diagnostic framework by capturing both neurophysiological and structural changes, allowing for earlier and more accurate detection of cognitive decline (20,25). Additionally, integrating EEG with functional MRI (fMRI) has proven valuable for patients with acute disorders of consciousness (DoC) in ICU settings, providing insights into residual consciousness that often go undetected with single-modality assessments (26). For sleep stage classification, multimodal EEG data incorporating EOG, and EMG data enhances the accuracy of classifying sleep stages like REM and NREM, supporting better diagnostic precision in sleep-related disorders (27,28).

These examples underscore the versatility and efficacy of multimodal EEG approaches across diverse clinical applications. By leveraging the complementary strengths of various physiological and neuroimaging modalities, multimodal EEG data provides a more comprehensive view of brain function and behavior, paving the way for more sensitive and specific diagnostics across a wide spectrum of neurological and psychiatric disorders (29).

### 1.5 Terminology

This review incorporates various terminologies and acronyms spanning machine learning models, human health data, and disease references. For machine learning and deep learning models, terms like SVM, KNN, CNNs, and LSTM are used to represent specific algorithms. Acronyms such as ECG, EEG, EMG, and EOG stand for Electrocardiography, Electroencephalography, Electromyography, and Electrooculography, respectively. Additionally, disease-related terms are included, such as AD for Alzheimer's disease and ASD for Autism Spectrum Disorder. A comprehensive list of these and other terms can be found in Appendix A.

### 1.6 Objective

This systematic review explores the state-of-the-art methods in ML and DL for processing EEG data in combination with other modalities. It examines many recent studies, offering a comprehensive overview for researchers experienced in traditional EEG data fusion or multimodal processing interested in leveraging ML and DL techniques. Additionally, the review aims to guide clinical researchers in applying ML and DL to multidimensional EEG data for specific clinical applications. The review provides detailed methodological insights into the various components of an ML-EEG pipeline, assisting readers in implementing these techniques in their work. Beyond highlighting trends and notable approaches, the review concludes with several recommendations to promote reproducible and efficient research in the field.

### 1.7 Organization

The review is structured as follows: Section [1](#) offers a brief introduction to essential concepts in EEG, ML, DL, and multimodal data, along with the objectives of the review. Section [2](#) explains the methodology of the systematic review, including the selection, assessment, and analysis of studies. Section [3](#) highlights the key analysis result of the selected studies, identifying key approaches in applying ML/DL in multimodal EEG data. Section [4](#) addresses critical issues and challenges in DL-EEG, providing recommendations for future research. Section [5](#) concludes with suggestions for future ML/DL multimodal EEG research directions.

**Paper Review Method**

### 2.1 Search methods

In our systematic review, we employed a structured approach to identify and analyse relevant literature on machine learning methods for multimodal EEG data in clinical applications. Our search strategy followed the PRISMA (Preferred Reporting Items for Systematic Reviews and Meta-Analyses) guidelines to ensure a comprehensive and unbiased review process (6).

We conducted our search across three major databases: PubMed, Web of Science, and Google Scholar. These databases were chosen due to their extensive scientific and clinical research papers repository.

For PubMed, we used a comprehensive search query incorporating key terms related to EEG, BCI, and machine learning models:

*(EEG OR Electroencephalogra\* OR Brain-Computer Interface OR Brain Computer Interface OR BCI) AND (Resting State OR Resting-State OR Eye\* OR Motor Imagery OR Neck Movement) AND (Deep Learning OR Machine Learning OR Transformer OR Multi\* Kernel Learning OR Ensemble Learning OR CNN OR RNN OR ANN OR Neural Network OR SVM OR Clustering OR Transfer Learning) AND (Multidimensional Data OR multimodal data OR multimodal data fusion OR multimodal learning OR data integration OR heterogeneous data integration OR heterogeneous data fusion) Filters: Full text, from 2012 - 2024.*

A similar structure was used for the Web of Science database, with the addition of "TS=" before each search term group to accommodate its search syntax.

Google Scholar required a different approach as it does not support multiple "AND"/"OR" groups in the search query, only the 'OR' conjunction. We separated our search into four categories focusing on specific research areas: resting state, eye closed/open, motor imagery, and neck movement. For each category, we conducted individual searches. For example, we first searched for papers containing resting state EEG data and then connected them with all other keywords using the 'OR' conjunction. We retrieved the first 150 to 200 papers from Google Scholar that fit our study criteria. This process was repeated for the other three research areas until we had a comprehensive collection of papers.

As a result of these methods, we obtained 45 related papers from PubMed, 154 related papers from Web of Science, and 564 papers from Google Scholar.

We categorized our search terms into four main groups to capture a broad yet specific range of relevant studies. Fundamental terms such as EEG, Electroencephalography, and BCI formed the core of our search. These terms are pivotal to our research focus and ensure the retrieval of studies central to EEG and BCI technologies.

Specific areas of interest within EEG research were identified, including resting-state (eye open/closed), event-related desynchronization (ERD), neck movement, and motor imagery. These terms helped refine our search for studies that align closely with our research objectives.

We included a wide range of machine learning models such as deep learning, neural networks (ANN, CNN, RNN), support vector machine (SVM), and ensemble learning. This ensured that our search covered the diverse computational approaches applied in EEG data analysis. Terms related to the handling and integration of complex data types were also included. This encompassed methodologies for multimodal data fusion, heterogeneous data integration, and multimodal learning, which is crucial for our review of how different data sources are managed in EEG research. This strategic categorization and the selection of search terms ensured a comprehensive and focused retrieval of relevant literature for our systematic review, as illustrated in the "Search Term Combination" (Figure 1).

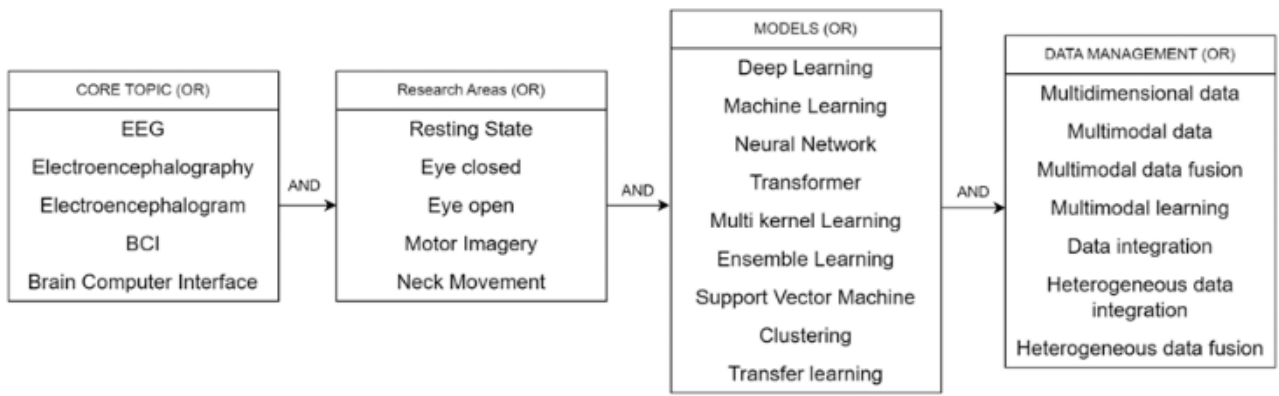

**Figure 1 - Search Term Combination**

*2.2 Filter methods*

Our filtering process was designed to systematically narrow down the vast number of search results to a manageable and relevant subset of papers. This multi-layered filtration process involved three main stages.

Initially, each review group member individually screened the retrieved papers' titles using our predefined search criteria. Papers whose titles clearly indicated relevance to our core topics, research areas, models, and data management techniques were included in this initial selection. This step ensured a diverse yet focused pool of initial results. The individually selected titles were then consolidated, and common selections were identified for further review.

The second stage involved a detailed review of the abstracts of the selected titles. This allowed us to evaluate the depth of content and its alignment with our research objectives. Abstracts providing insights into the study's methodology, objectives, results, and conclusions were considered for further review. Group discussions followed individual assessments to reach a consensus on the papers advancing to the final review stage.

The final stage involved an in-depth examination of the full texts of the shortlisted papers. This comprehensive review ensured that the methodologies, findings, and discussions were directly relevant to our research aims and provided valuable insights or models applicable to our study. Each paper was reviewed individually and then discussed in group sessions to finalize the selection of papers included in our systematic review. The process is illustrated in the " Research Paper Review Strategy Flow Chart " (Figure 2).

After this thorough paper review and filtering process, we decided to keep 16 papers covering clinical populations as our study's focus. Additionally, we retained 47 papers that only included healthy control EEG data for reference. In the later sections of this paper, we will analyse these 16 clinically related papers, creating summary tables that detail the preprocessing methods, data fusion methods, and machine learning applications used. This structured approach ensures that our review is both comprehensive and focused, enabling us to draw meaningful insights from the existing literature.

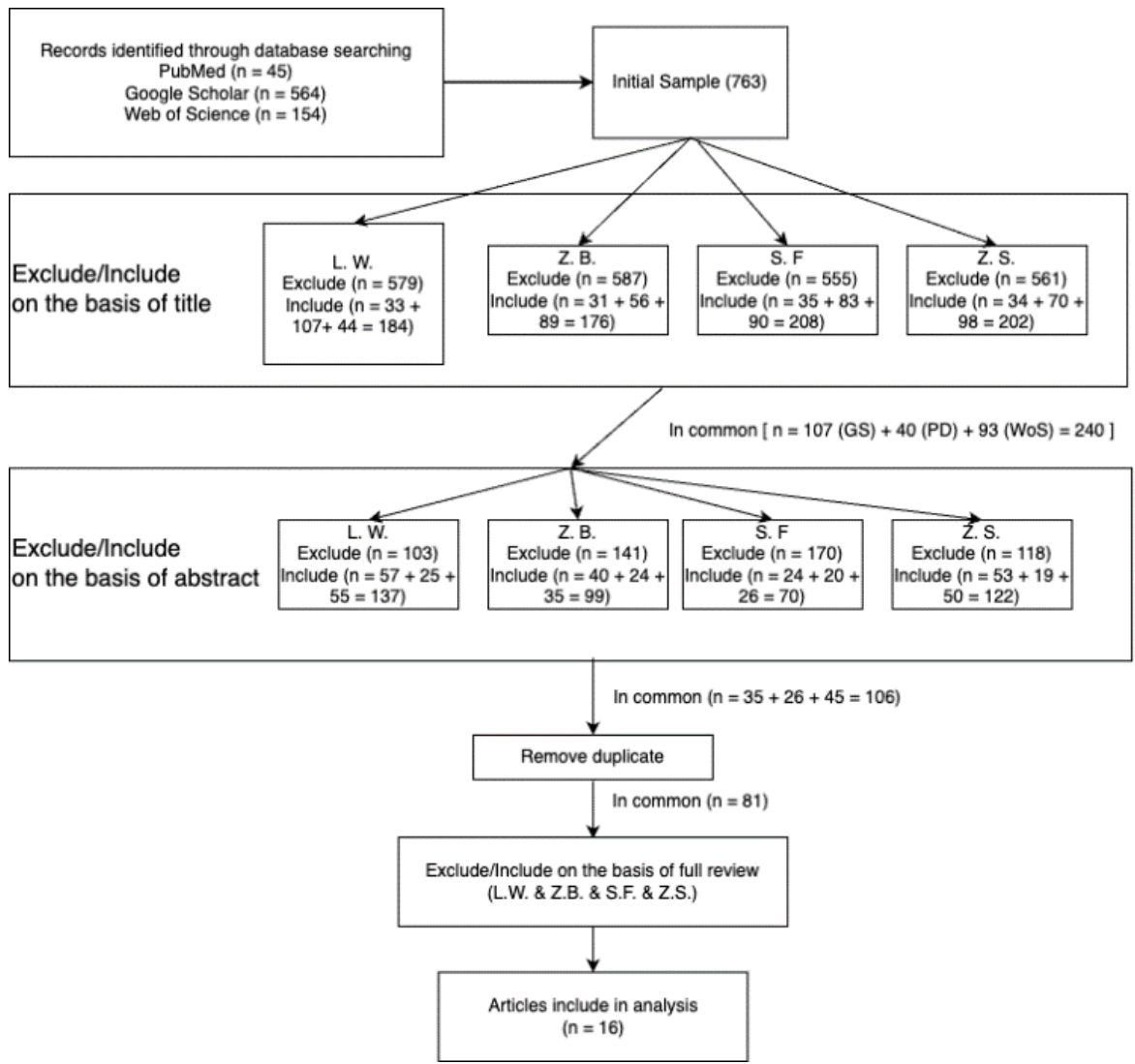

**Figure 2 - Research Paper Review Strategy Flow Chart.**

## Result

### 3.1 Which EEG multidimensional diseases/tasks have been explored with machine learning

#### 3.1.1 Neuropsychiatric Disorders

Multimodal learning has been applied to different neuropyshciatric disorders. For instance, a virtual reality biofeedback exposure therapy system for public speaking anxiety, recognition of mild depression, and prediciton of internet gaming disorder have all benefited from mutlimodal learning. Heart rate variability and EEG data recorded during virtual reality simulated public speaking enhanced the accuracy of detecting high arousal states (18). Predictions of internet gaming disorder were improved by training a multi-kernel SVM that combines, positron emission topography, EEG, and answers to an internet gaming disorder questionaire (30). Lastly, the detction of mild depression was improved by combining synchronized EEG and eye movement recordings during free viewing of neutral and negative faces through a multimodal deep learning model with fusion at the feature level (31).

#### 3.1.2 Neurological Disorders

Mild cognitive impairment (MCI), Alzheimer's disease, epilepsy, disorders of conciousness, and amyotrophic lateral sclerosis (ALS) have also benefited from mutlimodal learning. Early detection of MCI was enhanced by incorporating multimodal features including EEG, handwriting (32), speech and parameters derived during a virtual reality-based scene description task (25). Interestingly in Alzheimer's disease detection, combining EEG with cognitive questionairres does not improve the classificaiton accuracy of a multimodal model compared to classification with cognitive quesitonairres alone. Furthermore, a classifier traied on magnetic resonance imaging (MRI) alone outperformed all other unimodal and multimodal models in classifying Alzheimer's disease (20). Multimodal learning has been applied to epilepsy for both seizure detection and identification of epileptogenic networks. Seizure detection and identifying epileptogenic networks have been enhanced by combining funcitonal magnetic resonance imaging (fMRI) and EEG (33). Seizure detection has been improved by combining EEG and funcitonal near infarred spectroscopy (fNIRS) (22). Further, seizure freedom following mesieal temporal surgery was accurartely predicted by combining structural MRI, pre-operative tests, and intarcranial EEG features (34). More complex models have also had improved success in seizure deteciton by combining mutliple modalities of physiological data. Liu et al., enhanced seizure detection using a CNN by combining EEG, electrocardiography, and respiratory features (21). Zhang et al., enhanced tonic-clonic seizure detection, a high predictor of unexpected death in epilepsy, by combining EEG and EMG data to train a cyclic transformer. The cyclic transformer then only used EEG data to detect tonic-clonic seizures (23). Combining fMRI and EEG enhanced the accuracy of prognostic outcomes from individuals with acute disorders of conciousness (26). EEG, fNIRS, and data from a visual oddball task and a mathematical task were combined to accurately identify individuals with ALS (29).

### 3.1.3 Neurodevelopment Disorders

Multimodal machine learning models have been trained to accurately diagnose children with Autism Spectrum Disorder (ASD). Liao et al., developed a high performing model for diagnosing ASD while also identifying robust ASD markers for healthcare professionals. They used features extracted from EEG, eye-tracking, and facial expressions recorded during presentation of social and non-social videos (17). Other work has fused EEG and eye tracking data to achieve an overall accuracy of 87.5% for early detectyion of ASD (24).

### 3.1.4 Sleep stage classification

CNNs have been used for multimodal classification of sleepstages. Abdollahapour et al., combined EEG and EOG along with transfer learning across two different datasets (Sleep-EDF and Sleep-EDFx) to classify different sleep stages. Sleep casetes and sleep telemetry data was used from both datasets (28). Ellis et al., combined EEG, EOG, and chin EMG data to classify different sleep stages through sleep telemetry data from approximately 9 hours of sleep. The authors used a novel "ablation" approach to identify which modality was most important to classification accuracy. Interestingly, chin EMG had the least impact on model performance, whereas EEG was important for most stages and EOG was important for rapid eye movement and non-rapid eye movement stages (27).

### 3.2 Data fusion methods for multimodal EEG data

Multimodal Data fusion methods are commonly used to combine information from different modalities or signals so that a uniform input format can be built for the machine learning models. The fusion methods can also be used to obtain the more useful information while excluding the less important information for the model performance. Reducing information redundancy can potentially prevent the machine learning model from overfitting. In general, the data fusion methods overcome the limitations of individual modalities by incorporating complementary information from other modalities (29). There are three main levels of data fusion methods, including signal level, feature level, and decision level.

### 3.2.1 Signal Level

The signal-level data fusion methods combine the data from different signal sources, which are the modalities of the studies after preprocessing the multi-modal data. The fusion could optionally include the application of a modality selection method to choose among data of multiple modalities based on their importance on the machine learning model's performance. The fused data will be used for further feature extraction or directly as inputs to the machine learning model. For instance, in the study of Ellis et al. (27), preprocessed EEG, EOG, and EMG data have been fused together as input to a deep learning model with 1-D CNN architecture. An ablation approach has been applied in the study, which replaces the signal data of each of the 3 modalities that have been mentioned with a sinusoidal and Gaussian noise that can be commonly found in electrophysiology data to determine the importance of each modality on the performance of the CNN model.

Signal-level data fusion has only been employed in 2 (out of 15) clinical studies that have been full-text reviewed. It is because most of the clinical tasks needed to use the feature vectors, which are acquired through feature extraction and feature-level fusion methods, as inputs to the machine learning models, and the effects of modality selection methods on the signal level can be achieved by excluding all the features that are extracted from signal data of the corresponding modality during feature-level data fusion.

### 3.2.2 Feature Level

The feature-level data fusion refers to the process of combining the features extracted from different modalities, such as EEG signals and EOG signals. It is performed to create a more informative feature set. The combined features will be the input to the machine learning models.

There are various feature-level data fusion methods depending on the types of features and the machine learning models which the features' data will be fed into. Most conventional machine learning models (e.g., Regression, K-Nearest Neighbour, SVM, etc.) and deep learning models, such as the regular fully connected artificial neural network, take one-dimensional vectors of numeric values as inputs. Hence, intuitively, concatenating the values of the selected features into a 1-D feature vector has been a commonly used feature-level fusion method by many of the studies that have been reviewed. For example, in the study of Zhu et al. (31)They directly concatenated features from the EEG signals of each band and the EM features selected using the Best First approach. The combined features were then used as the input for the autoencoder. The feature-level data fusion process may also involve data transformation, transforming the original features into new features that can be used as inputs to the machine/deep learning model. For instance, in the study of Abdollahpour et al. (28), they transformed the EEG and EOG feature vectors into Horizontal Visibility Graphs (HVG), where each EEG and EOG feature is a node in the graph. They used a framework to draw HVGs on 2D Euclidean space as 2D images with a specific structure. Then, the images were fed to a 2D CNN model as inputs.

It is also worth mentioning that other than the step of utilizing data fusion methods, the overall process of feature-level data fusion usually also includes a feature selection step. The purpose of feature selection is to select the features important to model performance while excluding the ones that are not. By excluding the unimportant features, the number of parameters and, hence, the complexity of the machine learning model can be reduced so that the model will be less prone to overfitting. The feature-selection methods that have been employed vary a lot across different studies. Some of the techniques that have been used in multiple studies that we have reviewed are:

a. Selecting the features with the mRMR technique to achieve maximum relevance and minimum redundancy in the input feature vector. (24,29)

b. Cross-validation technique is used to examine models' performance with different combinations of features as inputs, retain the best-performing features, exclude unimportant features, and select the optimal number of features to be included in the input vector. (25,26,29)

*3.2.3 Decision Level*

The decision-level data fusion takes place when the training data is split into different sample and/or feature subsets, where each subset will be used as a training set of a separate machine learning model. Each trained model will generate a separate prediction result, then the decision-level data fusion method will be employed to merge the models' results (e.g., by a weight voting strategy) to generate the final result. Because decision-level data fusion is closely related to the concept of ensemble learning in the field of machine/deep learning, specific examples will be discussed in the next section (30,32). Decision-level data fusion has been applied in 5 (out of 15) clinical studies that have been full-text reviewed.

*3.3 Machine Learning Models Application*

This section provides an overview of the trends and advancements in machine learning models applied to multimodal EEG data in clinical studies. By examining the methods and outcomes from various research papers, we aim to highlight the most prevalent models, their applications, the evolution of techniques, and the performance metrics that underline the efficacy of these models in clinical contexts.

*3.3.1 Commonly used machine learning models*

The clinical application of machine learning models to multimodal EEG data has seen a variety of approaches, reflecting the diversity of tasks and datasets involved. Among the most popular models are SVM, Decision Trees (DT), K-Nearest Neighbors (KNN), CNNs, and RNNs, particularly LSTM networks.

SVMs are widely utilized across various clinical fields, including study of ASD detection, disorders of consciousness (DoC), mild depression, anxiety, and sleep stages classification. They have proven effective in integrating multimodal inputs such as EEG, fNIRS, PET scans, and speech data. Advanced variants like Radial basis function kernel and Multiple-Kernel SVMs further enhance their applicability in complex tasks such as mild cognitive impairment (MCI) screening and Internet Gaming Disorder (IGD) prediction (17,18,23–26,28–30,32,35).

DT are widely employed in clinical studies due to their simplicity, interpretability, and effectiveness. They are used in tasks such as anxiety detection, MCI screening, and sleep stage classification by integrating multimodal inputs like EEG, heart rate variability, and speech data. Random Forests (RF), an ensemble method based on Decision Trees, extend these capabilities by improving robustness and accuracy. RF has been applied in ASD detection, mild depression recognition, and predicting functional outcomes in disorders of consciousness, effectively handling complex multimodal datasets such as fMRI, eye movement, and handwriting dynamics (17,18,25,26,28,32,35).

CNNs are extensively utilized in clinical studies for their ability to extract hierarchical spatial features from complex data. They are particularly effective in tasks such as epileptic seizure detection, where EEG is combined with respiratory signals and ECG to enhance predictive accuracy (21). CNNs are also applied in explainable sleep stage classification, leveraging multimodal data such as EEG, EOG, and EMG to provide interpretable insights into classification outcomes (27). Additionally, CNNs are used in multimodal epilepsy analysis, integrating EEG with resting-state fMRI to capture both temporal and spatial features, showcasing their versatility in handling diverse multimodal datasets (33).

KNN is a simple yet effective machine learning model applied in various clinical tasks involving multimodal EEG data. In Autism Spectrum Disorder (ASD) detection, KNN is utilized to classify children as either ASD or typically developing based on EEG and features like facial expressions, eye fixation, and eye-tracking data (17,24). For anxiety detection, KNN integrates EEG with heart rate variability (HRV) to analyze physiological arousal states (18). Additionally, in sleep stage classification, KNN processes EEG and horizontal EOG data, demonstrating its adaptability in handling multimodal inputs across diverse clinical applications (28).

RNNs, particularly LSTM networks, are widely used in multimodal EEG studies for their ability to model temporal dependencies in sequential data. LSTMs are applied in seizure detection and epilepsy monitoring, integrating EEG with fNIRS to capture dynamic patterns across modalities (22). In epilepsy-related tasks, LSTMs are further combined with resting-state fMRI data, enabling the identification of epileptogenic networks and seizure foci by leveraging the temporal and spatial features of multimodal inputs (33). Hybrid models like ResNet-LSTM enhance these capabilities by integrating residual networks for feature extraction with LSTMs for temporal modeling, proving effective in detecting tonic-clonic seizures and predicting SUDEP (23).

In addition to these widely used models, several other machine learning models have been applied to multimodal EEG data, showcasing their adaptability to diverse clinical tasks. For anxiety detection, models like Gaussian Naïve Bayes (GNB), quadratic discriminant analysis (QDA), AdaBoost (ADB), and multilayer perceptron (MLP) effectively integrate EEG with heart rate variability (HRV) data (18). Advanced architectures such as Cyclic Transformers (CT) and their multimodal extensions (MICT), along with ChronoNet, are employed in tonic-clonic seizure detection, utilizing multimodal inputs like accelerometer (ACC) and electromyography (EMG) (23). Gradient Boosting Decision Trees (GBD Tree), Self-Normalizing Neural Networks (SNN), and Batch Normalized Multilayer Perceptron (BNMLP) are applied in mild depression recognition using EEG and eye movement data (35). Additionally, XGBoost demonstrates strong performance in diagnosing mild cognitive impairment (MCI) using handwriting dynamics, speech data, and digitized cognitive parameters, highlighting its versatility across neurological and psychiatric applications (25,32). These diverse models contribute to enhancing the accuracy and robustness of clinical predictions in multimodal EEG studies.

In conclusion, a wide array of machine learning models has been applied to multimodal EEG data. Each model's unique strengths and capabilities contribute to more accurate and robust clinical predictions, demonstrating the value of leveraging various machine learning approaches to tackle complex neurological and psychiatric conditions.

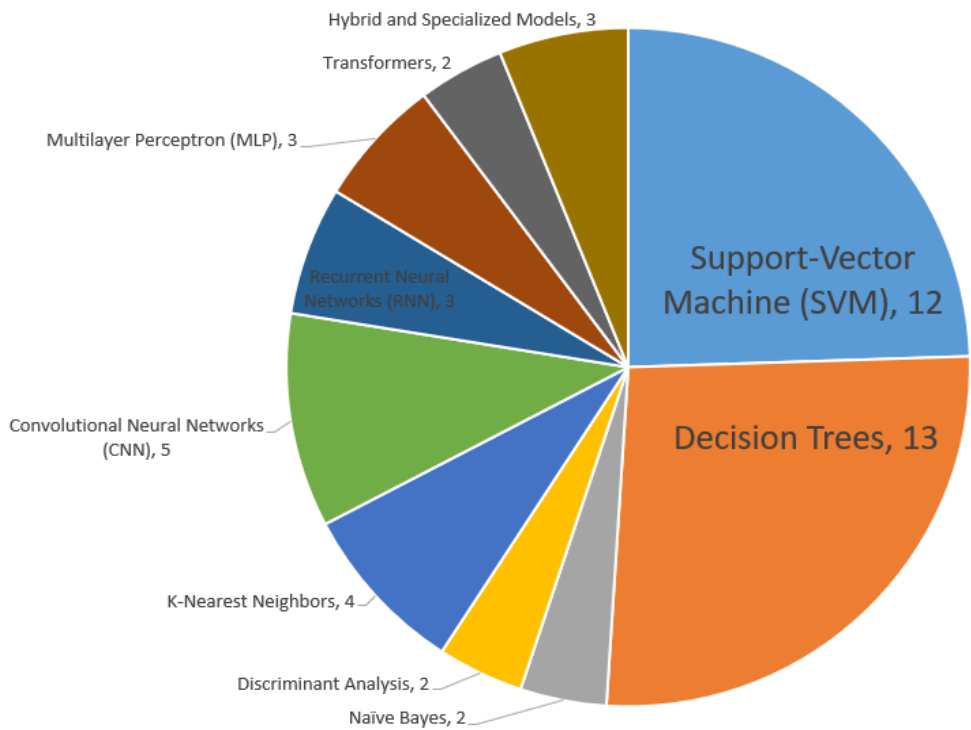

Figure 3 - Frequency of Machine Learning Models Used in Clinical Studies

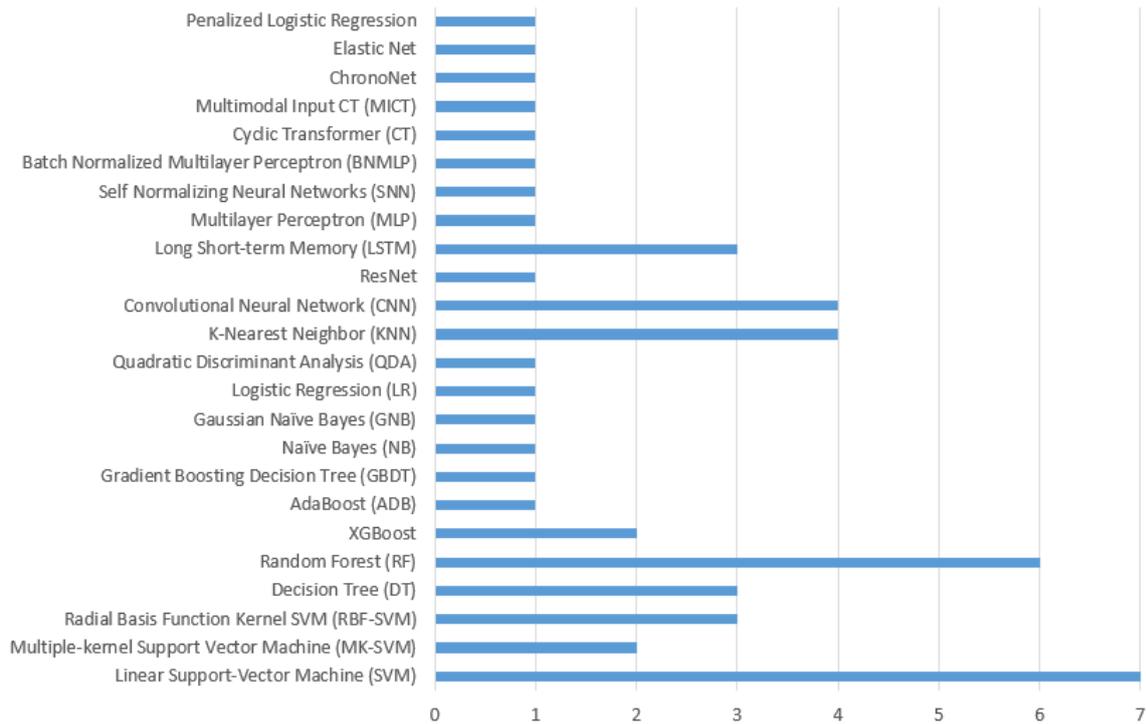

Figure 4 – Machine Learning Model Counts Used on Clinical Studys

### 3.3.2 Diseases/Task specific application

The application of machine learning models to multimodal EEG data in clinical settings exhibits distinct trends based on the specific clinical tasks addressed. These trends highlight the versatility and adaptability of machine learning techniques in tackling a wide range of neurological and psychiatric conditions.

| Study | Task | Data | Best ML Model | Single Modality Accuracy (EEG only) | Multi-modality Result Accuracy |
|---|---|---|---|---|---|
| (17) | ASD | EEG Facial Expression Eye-tracking data | Random Forest on Single Modality; Weighted Naive Bayes, K-means, VGG16 On Multi-modality | 83.7 | 87.5 |
| (24) | | EEG Eye-tracking data Computational Data | SVM | 84 | 92 |

Table 1 – Machine Learning Method and Accuracy on ASD task

One prominent area is the detection and classification of ASD. Studies in this domain often integrate EEG data with other modalities, such as eye fixation and facial expression data, to improve diagnostic accuracy. Machine learning models such as SVM, RF, and KNN are frequently employed, with multimodal approaches consistently outperforming unimodal ones. For instance, integrating EEG with eye fixation and facial expression data significantly improved classification accuracy, demonstrating the effectiveness of multimodal data fusion in enhancing model performance for ASD detection(17,24).

As shown in Table 1, these studies typically integrate EEG data with additional modalities like facial expressions and eye-tracking data. For ASD detection, a Random Forest model achieved an accuracy of 83.7% with EEG data alone. However, the integration of multimodal data, including facial expression and eye-tracking information, increased the accuracy to 87.5% using a combination of models such as Weighted Naive Bayes, K-means, and VGG16. (17)

In another example, using a Support Vector Machine (SVM) with EEG and eye-tracking data, the accuracy improved from 84% (single-modality EEG) to 92% with multimodal data. This comparison underscores the effectiveness of incorporating multiple data sources, as it consistently enhances diagnostic accuracy, demonstrating the advantages of multimodal data fusion in machine learning for ASD detection.(24)

| Study | Task | Data | Best ML Model | Single Modality Accuracy (EEG only) | Multi-modality Result Accuracy |
|---|---|---|---|---|---|
| (21) | Epileptic Seizure Detection | EEG ECG Respiratory signal | CNNs (Con1D) | 57.67 | 61.06 |
| (22) | | EEG fNIRS | RNNs LSTM | 97.6 ± 0.4 SD % | 98.3 ± 0.8 SD % |
| (33) | | EEG rs-fNIRs | CNNs for EEG feature Extraction; LSTM for multimodal integration of rs-fMRI and EEG; SVM | NA | Accuracy: 96.66% Precision: 97.57% Sensitivity: 95.83% |
| (23) | | EEG ACC EMG ECG | CT MICT | Sensitivity: 100%, F1: 44.2% | Sensitivity: 100%, F1: 60.9% |

Table 2 – Machine Learning Method and Accuracy on Epileptic Seizure Detection task

In the realm of seizure detection and epilepsy monitoring, the combination of EEG with other physiological signals, such as fNIRS, respiratory signals, and ECG, has proven highly effective. Models like CNNs, LSTM networks, and hybrid architectures are commonly used to leverage the rich information from these multimodal datasets. These approaches have shown remarkable accuracy in detecting epileptic seizures and predicting surgical outcomes in epilepsy patients (21,22,33).

From this perspective, deep learning methods, particularly those involving neural networks, are frequently applied to epileptic seizure detection in clinical populations. Liu et al. (2020) proposed a method for seizure detection using a convolutional neural network (CNN) that integrates multiple biosignals, including EEG, ECG, and respiratory signals. Their study demonstrated that models trained on multimodal data, specifically using a CNN model with a Conv1D layer, outperformed models using only EEG data, with accuracy improving from 57.57% to 61.06%. Similarly, Sirpal et al. (2019) found that incorporating fNIRS data also enhanced seizure detection accuracy. By using both EEG and fNIRS data, accuracy increased from 97.6% ± 0.4 SD to 98.3% ± 0.8 SD.

Although Hosseini et al. (2020) did not directly compare multimodal data with EEG-only data, their study demonstrates that using multimodal data can achieve high accuracy in epileptic seizure detection. They applied deep learning and edge computing techniques on epileptic EEG and rs-fMRI data, achieving impressive results across various metrics: accuracy of 96.66%, precision of 97.57%, and sensitivity of 95.83%, all reflecting a high level of detection accuracy.(33) Zhang et al. (2024) utilized the latest transformer models and techniques to enhance the accuracy of seizure detection by integrating multimodal data, including EEG, ACC, EMG, and ECG signals. By employing Multimodal Input Cyclic Transformers, they achieved a significant improvement in performance metrics. Notably, while maintaining a sensitivity of 100%, they increased the F1 score from 44.2% to 60.9%, highlighting the robustness of their approach in detecting tonic-clonic seizures. Additional details are provided in Table 2.

| Study | Task | Data | Best ML Model | Performance |
|---|---|---|---|---|
| (27) | Sleeping stage classification on clinical population | EEG EOG EMG | CNNs | Highest precision and F1 scores on NREM2 class Precision: 79.35±03.92 Recall: 68.71±08.51 F1-score: 73.28±04.76 |
| (28) | | EEG EOG | CNNs (trained with transfer learning) | Accuracy: 94.34% |

Table 3 - Machine Learning Method and Accuracy on Sleeping Stage Classification Task

Multimodal data from EEG and EOG are often utilized to classify sleep stages. Advanced models like CNNs, especially those employing transfer learning, have demonstrated high accuracy in distinguishing between different sleep stages. This trend underscores the importance of combining spatial and temporal features to capture the nuances of sleep patterns effectively(27,28).

Although both studies(27,28) employed CNN models and integrated EOG data with EEG data, neither provided a comparison of the effects of using EEG and EOG together versus using EEG alone. Reviewing studies (17,21–24) on other diseases or tasks on clinical population that utilize multimodal data and comparing their outcomes with those relying solely on EEG could enhance the comprehensiveness and validity of the findings

| Study | Task | Data | Best ML Model | Performance |
|---|---|---|---|---|
| (18) | Neuropsychiatric Disorders (Anxiety, Mild Depression, Internet Gaming Disorder) | EEG GSR HR EMG | RF | Accuracy: 87.5% |
| (31) | | EEG Eye-tracking Data | Linear SVM | Accuracy: 81.88 ± 0.19 |
| (30) | | PET CF | Multiple-kernel SVM | Optimal Case: Weights: PET 0.32, CF 0.62, EEG 0.06 Accuracy: 84.6% Sensitivity: 89.3% Specificity: 79.2%. |

Table 4 - Machine Learning Method and Accuracy on Neuropsychiatric Disorders task

Mental health applications, such as anxiety detection and mild depression recognition, benefit from the integration of EEG with other biosignals like galvanic skin response (GSR), heart rate (HR), and EMG. Machine learning models including GNB, QDA, SVM, and RF have been applied to classify emotional states and detect mental health conditions. The multimodal approach in these studies enhances the sensitivity and specificity of the models, providing a more comprehensive understanding of the subject's mental state (18,31).

From Table 4, we can see that for the neuropsychiatric disorders task, due to the different types of diseases, there are Anxiety, Mild Depression, Internet Gaming Disorder, so the data used to fuse with EEG data are also different, including GSR, HR, EMG, Eye-tracking Data, PET and CF. They all used a variety of machine learning algorithms or models, but the best results were obtained with SVM or RF. Like the sleep stage analysis studies, none of these studies provided a comparison with the results of modeling using only EEG data.

| Study | Task | Data | Best ML Model | Single Modality Accuracy (EEG only) | Multi-modality Result Accuracy |
|---|---|---|---|---|---|
| (25) | Cognitive Disorders (A precursor to Alzheimer's disease, Mild Cognitive Impairment, Amnestic MCI) | EEG SD DCP | SVM | 83 | 87 |
| (32) | | EEG Handwriting | SVM (RBF kernel) | 86.1 | 96.73 |
| (20) | | EEG EOG sMRI MMSE Score | Penalized logistic regression with Elastic Net regularization | AUC: 0.76 | AUC: 0.98 |

Table 5 - Machine Learning Method and Accuracy on Cognitive Disorders Task

In the assessment of MCI and Alzheimer's disease (AD), combining EEG with structural MRI, speech data, and handwriting dynamics has shown significant promise. Models such as SVM, RF, and XGBoost are used to analyse these diverse data sources, improving the early detection and monitoring of cognitive decline. The integration of multiple data types helps in capturing a more complete picture of cognitive function, leading to more accurate and reliable diagnostic tools (20,25,32).

| Study | Task | Data | Best ML Model | Single Modality Accuracy (EEG only) | Multi-modality Result Accuracy |
|---|---|---|---|---|---|
| (26) | Neurological Conditions (Long-term functional outcomes of patients with acute disorders of consciousness, Classification between Amyotrophic Lateral Sclerosis (ALS) patients and healthy controls) | EEG fMRI | Random Forest | AUC(Synek): 0.67 (95% CI: 0.65-0.70) | AUC(Synek + fMRI): 0.75 |
| (29) | | EEG fNIRS | SVM | 73.23 | 85.38 |

Table 6 - Machine Learning Method and Accuracy on Neurological Conditions Task

Table 6 summarizes the machine learning method and accuracies for neurological condition-related tasks using EEG and other multimodal data sources. For the task of predicting long-term functional outcomes in acute disorders of consciousness (DoC), the best-performing model was a Random Forest-based EEG Multi-feature Model V, which integrated EEG features Synek scoring, ABCD classification, and EEG markers derived from resting-state EEG segments, highlighting the efficiency of EEG based assessments for both short- and long-term prognostication in DoC patients.

For the classification task between ALS patients and healthy controls, a multi-modality SVM model that combined EEG and fNIRS data demonstrated a high accuracy of 85.38%, compared to 73.23% accuracy with EEG alone. This indicates that integrating multiple data modalities significantly enhances diagnostic accuracy and reliability in neurological conditions classification tasks (26,29).

Overall, the trend toward using multimodal data in combination with sophisticated machine learning models reflects a growing recognition of the value of integrating diverse physiological and behavioral signals. This approach enhances the performance of diagnostic and predictive models and provides deeper insights into the underlying mechanisms of various neurological and psychiatric disorders.

*3.3.3 Machine Learning Model Architecture Design*

The design of machine learning models for multimodal EEG data in clinical applications involves various architectures and data fusion techniques to optimize performance and accuracy. This section reviews the model architectures and the decision-level data fusion methods employed in the reviewed studies.

The architectural design of machine learning models applied to multimodal EEG data in clinical studies showcases a wide range of approaches tailored to specific tasks and data characteristics. CNNs are widely used in the analysis of multimodal EEG data due to their robust feature extraction capabilities. In epileptic seizure detection, CNNs demonstrate significant adaptability by leveraging both Conv1D and Conv2D architectures, tailored to the nature of the input data. Conv1D processes sequential one-dimensional signals, while Conv2D handles two-dimensional representations, such as combined EEG, ECG, and respiratory signals, capturing spatial relationships among them. These two approaches enable the detection of intricate patterns and improves classification outcomes by fusing diverse biosignals (21). Additionally, CNNs have been employed in automated sleep staging, where one-dimensional convolutional layers, max pooling, and dropout layers are combined to refine feature extraction and enhance classification accuracy. These architectures are often supported by optimization techniques like transfer learning, enabling the application of pre-trained models to multimodal datasets, which further enhances their performance (27,28). The versatility and hierarchical learning capabilities of CNNs make them particularly effective for tasks involving spatial and temporal data, solidifying their role in multimodal EEG analyses.

RNNs, particularly LSTM networks, are widely utilized for their ability to model sequential and time-series data, making them well-suited for analyzing multimodal EEG data in epilepsy-related studies. In seizure detection and epilepsy monitoring tasks, LSTM networks have been employed to integrate EEG data with functional near-infrared spectroscopy (fNIRS), leveraging the temporal dependencies between these modalities to enhance prediction accuracy. These architectures incorporate hyperbolic tangent activation for LSTM units, logistic sigmoid for gates, and categorical cross-entropy loss functions optimized with Adam, effectively capturing both long-term and short-term dependencies in the data (22). In multimodal integration tasks, LSTM models are utilized to combine rs-fMRI and EEG data, enabling the identification of epileptogenic networks and seizure foci. This approach benefits from the distinct temporal and spatial features of each modality, with LSTM layers processing sequential data to predict seizure intervals. The combination of feature extraction from convolutional layers and multimodal integration through LSTM networks demonstrates the effectiveness of these architectures in managing complex multimodal datasets (33). Advanced hybrid models like ResNet-LSTM and ChronoNet extend the capabilities of LSTM by combining residual networks or other advanced features with temporal modeling. These architectures, benchmarked against cyclic transformers and multimodal input cyclic transformers in recent studies, exhibit strong performance in tonic-clonic seizure detection and other epilepsy-related applications. Their ability to balance feature extraction, temporal modeling, and multimodal integration makes RNN-LSTM models a powerful choice for handling EEG data in clinical contexts(23,36,37).

Support Vector Machines (SVMs) are extensively utilized across a wide range of clinical tasks involving EEG data, demonstrating their versatility and robustness. In autism spectrum disorder (ASD) detection, SVMs are applied to multimodal data such as facial expression and eye fixation features, effectively distinguishing children with ASD from typically developing individuals (17,24). For biofeedback-driven self-guided virtual reality exposure therapy, SVMs process EEG data alongside heart rate variability (HRV) to classify arousal states, showcasing their adaptability in combining physiological and behavioral features (18). SVMs also play a pivotal role in predicting outcomes and classifying neurological disorders. In mild cognitive impairment (MCI) studies, Radial Basis Function kernel SVMs (RBF-SVMs) are employed to integrate speech data and cognitive parameters, offering enhanced accuracy in detecting this precursor to Alzheimer's disease (25,32). Similarly, in the classification of patients with pathological conditions versus healthy controls, non-linear polynomial kernel SVMs leverage hybrid EEG and fNIRS data, utilizing nested cross-validation to optimize feature selection and performance (29). Linear SVMs are employed in studies predicting long-term functional outcomes in patients with disorders of consciousness, as well as in depression recognition tasks involving EEG and eye movement data (26,31). Advanced variants of SVMs, such as Multiple-Kernel Support Vector Machines (MK-SVMs), further enhance performance by integrating multiple data sources like PET scans, EEG, and clinical features, as seen in Internet Gaming Disorder prediction (30). Feature-fusion approaches with SVMs are also applied in tonic-clonic seizure detection, integrating EEG, accelerometer, and electromyography data to refine classification results (23). In sleep stage classification, SVMs complement deep learning models by processing features derived from EEG and horizontal EOG signals, highlighting their flexibility in supporting multimodal data analysis (28). The frequent application of SVMs in EEG-based studies underscores their efficacy in handling high-dimensional and multimodal datasets. Their adaptability to various kernels and the ability to fuse diverse data types make them a cornerstone in machine learning applications for clinical EEG analysis.

Transformers have emerged as powerful tools for analyzing multimodal EEG data, particularly in tasks requiring integration of multiple modalities and robust feature extraction. In tonic-clonic seizure detection and the prediction of sudden unexpected death in epilepsy (SUDEP), the Cyclic Transformer (CT) demonstrates a novel architecture that extends beyond conventional transformer applications. The CT employs an encoder-decoder-reconstructor design, integrating a classifier within its architecture to support direct classification tasks. Unlike traditional methodologies focused on reconstructing missing modalities, the cyclic transformer employs a circular modality translation mechanism, enabling it to learn intermediate representations between source and target modalities. This design not only improves the representation learning process but also effectively handles incomplete datasets, addressing challenges common in biomedical signal processing. The Cyclic Transformer can be further extended into the Multimodal Input Cyclic Transformer (MICT), which processes two modalities simultaneously. In this architecture, one modality serves as the primary input, while the other acts as a supporting modality. Each modality is processed independently through transformer blocks to enhance its representation, followed by cross-attention mechanisms that allow the supporting modality to selectively enhance the feature extraction of the primary modality. This extension demonstrates the versatility of transformer-based architectures in leveraging intermodal relationships to improve classification performance (23,36,37).

In summary, the reviewed studies demonstrate a diverse range of machine learning architectures tailored to multimodal EEG data analysis. CNNs excel in capturing spatial hierarchies and are frequently used in tasks like seizure detection and sleep stage classification. RNNs, particularly LSTM networks, effectively handle temporal dependencies in sequential data, making them ideal for integrating multimodal inputs such as EEG and fNIRS. SVMs are robust classifiers widely applied across various clinical tasks, leveraging different kernel functions to adapt to linear and non-linear data distributions. Advanced transformer-based models, such as the CT and MICT, showcase their ability to process complex multimodal data, offering superior performance in scenarios involving incomplete datasets. These models, along with hybrid architectures like ResNet-LSTM, highlight the progression towards sophisticated designs that integrate feature extraction, temporal modeling, and intermodal relationships, addressing the unique challenges of clinical EEG data analysis.

*3.3.4 Decision-Level Data Fusion*

Decision-level data fusion is a pivotal strategy in enhancing the performance of machine learning models by integrating the outputs of multiple classifiers or models to reach a final, consolidated prediction. This approach leverages the strengths of various models, thereby improving the overall accuracy and robustness of clinical applications.

One exemplary application of decision-level data fusion is observed in the detection of ASD, where a hybrid fusion framework is employed. This framework combines feature-level and decision-level fusion, utilizing a weighted Naive Bayes algorithm to integrate sub-decisions from different feature sets. Specifically, behavioral features such as facial expressions and eye fixation data are fused at the feature level, while physiological features from EEG are processed separately. The sub-decisions derived from these features are then combined at the decision level using weighted coefficients that reflect the relative importance of each attribute, significantly enhancing classification accuracy (17)

In the context of detecting MCI, decision-level data fusion is implemented through a weighted voting strategy. This study integrates features from EEG, speech data, and digitized cognitive parameters. By employing a weighted voting strategy, the contributions of different modalities are balanced, improving the detection accuracy of MCI. This approach ensures that the final decision benefits from the complementary strengths of each modality, thereby providing a more holistic assessment of cognitive impairment (25). Similarly, a dual fusion strategy is utilized in the diagnosis of MCI based on handwriting dynamics and qEEG. Here, features extracted from both handwriting and EEG data are fused at the decision level. The final classification results are obtained by combining predictions from multiple base classifiers through a voting mechanism, which aggregates individual classifier decisions to enhance overall accuracy (32).

Another notable example is in predicting IGD, where the MK-SVM method is employed. This technique integrates multimodal data by modifying the kernel function to combine individual kernel matrices from each modality into a single kernel matrix. The MK-SVM model then uses this combined kernel matrix to perform classification, effectively handling the complexities of multimodal data and improving prediction accuracy (30).

Furthermore, in the integration of rs-fMRI and EEG data for seizure prediction, LSTM networks are used to capture temporal dependencies, and their outputs are combined with SVM classification results. This combination leverages the temporal analysis capabilities of LSTMs with the robust classification power of SVMs, resulting in enhanced prediction accuracy for seizure occurrences (33).

The implementation of decision-level data fusion in these studies underscores its critical role in clinical applications involving multimodal EEG data. By effectively integrating the outputs of diverse models, decision-level fusion boosts accuracy and ensures that the final predictions are more reliable and robust, thereby contributing to better clinical outcomes.

*3.3.5 Evolution of Techniques*

The evolution of machine learning techniques for multimodal EEG data in clinical applications is evident from the progression of methodologies and models over the years. By analyzing the publication dates of the reviewed papers, we can trace how the field has developed and adapted to new challenges and technological advancements.

In the late 2010s, machine learning techniques for multimodal EEG data analysis began to integrate traditional classifiers and deep learning approaches. RNN, particularly LSTM networks, were employed to combine EEG with functional near-infrared spectroscopy (fNIRS), effectively capturing temporal dependencies for seizure detection. Meanwhile, traditional models like linear SVMs, gradient boosting decision trees, and batch-normalized neural networks demonstrated their effectiveness in tasks such as depression recognition by leveraging multimodal inputs like EEG and eye movement data. These studies laid the groundwork for integrating multimodal features with diverse machine learning frameworks (22,35).

Entering the 2020s, advancements in neural network architectures, particularly CNNs, gained prominence for tasks like seizure detection and sleep stage classification. These models effectively captured spatial patterns within multimodal data, often integrating EEG with structural MRI, cognitive scores, and other physiological signals. Hybrid approaches also emerged, combining CNNs for feature extraction and LSTMs for temporal modeling, enabling accurate epilepsy predictions through multimodal inputs like rs-fMRI and EEG. Additionally, ensemble methods such as random forests and elastic net were utilized for integrating multimodal features in diagnosing conditions like Alzheimer's disease and amnestic mild cognitive impairment (aMCI), showcasing increasing sophistication in feature selection and model design (20,21,33).

By the early 2020s, the focus shifted toward explainability and robustness in machine learning applications. For example, explainable sleep staging models incorporated CNNs with dropout layers and class imbalance handling mechanisms to improve performance. SVMs with advanced kernels, such as non-linear polynomial kernels, were employed in ALS classification, leveraging EEG and fNIRS data through nested cross-validation and hybrid frameworks to enhance model accuracy and interpretability. These approaches emphasized the need for both precision and transparency in clinical applications (27,29).

As the mid-2020s approached, ensemble learning and kernel-based techniques became increasingly prevalent. Random forests, XGBoost, and multiple-kernel SVMs were utilized for diagnosing conditions like ASD, Internet Gaming Disorder (IGD), and mild cognitive impairment (MCI), effectively handling multimodal inputs such as EEG, PET scans, and eye-tracking data. These models demonstrated significant improvements in handling diverse data types and integrating information from multiple modalities, further advancing diagnostic capabilities (17,30).

More recently, advanced transformer architectures like the Cyclic Transformer (CT) and Multimodal Input Cyclic Transformer (MICT) have set new benchmarks in handling complex multimodal datasets. These transformers, capable of circular modality translation and cross-attention mechanisms, efficiently processed EEG, accelerometer, and electromyography data while addressing challenges like missing modalities. Compared to hybrid models like ResNet-LSTM and ChronoNet, transformers exhibited superior performance in capturing both temporal and spatial relationships within multimodal data, exemplifying the progression toward more flexible and powerful architectures (23,36,37).

The evolution of techniques over the years underscores a clear trajectory towards more sophisticated and integrated approaches. From basic linear models to advanced deep learning architectures, the progression highlights the continuous effort to improve machine learning models' accuracy, reliability, and interpretability for multimodal EEG data in clinical settings. This evolution demonstrates technological advancements and reflects a deeper understanding of the complexities involved in neurological and psychiatric disorders.

*3.3.6 Performance Metrics*

Performance metrics are crucial in evaluating the effectiveness of machine learning models in clinical applications. The highest accuracy reported in each reviewed paper indicates the best-performing model for specific clinical tasks. Additionally, the consistent observation across studies is that multimodal data integration invariably enhances model performance compared to unimodal data.

In the context of ASD detection, the integration of EEG with eye-tracking data and computational data achieved the highest accuracy. The SVM model, utilizing multimodal data, reached an impressive accuracy of 92%, significantly outperforming models that relied on single modalities (17,24).

For anxiety detection in biofeedback-driven therapy, the RF model again demonstrated superior performance. Combining EEG with GSR, HR, and EMG data, the model achieved an accuracy of over 85%, highlighting the value of incorporating multiple physiological signals (18).

In epileptic seizure detection, the use of CNN integrating EEG, respiratory signals, and ECG data showed substantial improvements. The Conv2D model attained an accuracy of 64.96%, illustrating the enhanced detection capabilities provided by multimodal data (21). Similarly, in studies involving EEG and fNIRS data, LSTM achieved an accuracy of 98.3%, further validating the effectiveness of multimodal approaches (22).

For tonic-clonic seizure detection, CT combined with EEG, accelerometer (ACC), and EMG data reached a sensitivity of 100% and an F1 score of 60.9%, showcasing the potential of advanced models in handling complex multimodal data(23).

In sleep stage classification, transfer learning with CNNs yielded the highest accuracy of 94.34%, significantly outperforming models trained without transfer learning and those using unimodal data (28). This underscores the advantage of leveraging pre-trained models and multimodal integration in improving classification accuracy.

The detection of MCI also benefited from multimodal data integration. SVM combined with handwriting dynamics and qEEG data achieved the highest accuracy of 96.73%, demonstrating the significant boost in performance when multiple data sources are utilized (32).

In the context of predicting IGD outcomes, an MK-SVM incorporating PET, EEG, and clinical features achieved an accuracy of 84.6%. This result highlights the importance of integrating diverse data types to enhance model robustness and predictive power (30).

Finally, explainable AI techniques applied to sleep stage classification using multimodal electrophysiology time-series data achieved notable precision and recall rates, further emphasizing the importance of multimodal data in improving model interpretability and performance (27).

Overall, the highest accuracies reported in these studies underscore the superior performance of machine learning models that utilize multimodal data compared to those relying on unimodal data. The consistent trend across various clinical tasks demonstrates that integrating multiple physiological and behavioral signals provides a more comprehensive understanding of the underlying conditions, leading to more accurate and reliable clinical predictions.

## Discussion

Choosing the appropriate machine learning model for EEG data analysis depends on the specific task, data characteristics, and desired outcomes. SVMs, particularly linear and RBF kernels, are ideal for classification tasks in high-dimensional data with small to medium-sized datasets due to their robustness against overfitting and efficiency in handling complex patterns. Decision Trees and their ensemble variants, such as Random Forests and XGBoost, are preferred when interpretability and handling of large datasets are crucial, offering insights into feature importance and enhancing model robustness through ensemble methods. CNNs are best suited for tasks that involve complex spatial hierarchies, such as detecting intricate patterns in EEG signals, making them highly effective for image and time-series data analysis. RNNs), including LSTMs, excel in analyzing sequential and time-series data due to their ability to capture long-term dependencies, making them suitable for tasks that involve temporal dynamics in EEG signals. Simpler models like KNN and MLP are effective for smaller datasets and adaptable to a variety of tasks, while advanced models like Transformers and specialized hybrid models are recommended for handling multimodal data and capturing complex relationships, providing tailored solutions for specific EEG analysis challenges.

## Conclusion

Multimodal EEG data has been effectively applied to ML and DL techniques to a variety of clinical tasks. Applications include the diagnosis of neuropsychiatric disorders, neurological conditions (e.g., seizure detection), neurodevelopmental disorders (e.g., ASD), and sleep stage classification. Multimodal EEG data has demonstrated its ability to enhance the accuracy of ML and DL models across these tasks. Different types of data are typically fused with EEG data depending on the clinical task. For instance, eye-tracking data is often combined with EEG to diagnose neurodevelopmental disorders such as ASD, while fNIRS and ECG are commonly integrated with EEG for detecting neurological conditions like seizures. Various ML and DL models have been applied to these tasks, with models such as SVM and Decision Trees frequently used in clinical applications. In contrast, DL models, including CNNs, RNNs, and LSTM networks, have been employed for more complex tasks. However, the use of Transformers for multimodal EEG data in clinical datasets remains limited, with only one study (out of 16 reviewed) exploring their potential. We recommend future research focus on leveraging advanced DL techniques, such as Transformers, to further improve performance in clinical tasks. Additionally, incorporating more diverse data types, when feasible, may enhance model accuracy and expand the scope of clinical applications. This approach has the potential to address a broader range of clinical challenges and contribute to the advancement of multimodal EEG-based research.

## Appendix A. List of Terminology

| | |
|---|---|
| ACC | Accelerometer |
| ADB | AdaBoost |
| AD | Alzheimer's disease |
| ASD | Autism Spectrum Disorder |
| BCI | Brain-computer Interface |
| CNNs | Convolutional Neural Networks |
| CT | Cyclic Transformers |
| DL | Deep Learning |
| DoC | Disorders of consciousness |
| DT | Decision Trees |
| ECG | Electrocardiography |
| EEG | Electroencephalography |
| EMG | Electromyography |
| EOG | Electrooculography |
| ERD | Event-related desynchronization |
| fMRI | functional magnetic resonance imaging |
| fNIRS | functional near-infrared spectroscopy |
| GBDT | Gradient Boosting Decision Trees |
| GNB | Gaussian Naïve Bayes |
| GSR | galvanic skin response |
| HR | Heart Rate |
| HVG | Horizontal Visibility Graph |
| IGD | Internet Gaming Disorder |
| KNN | K-Nearest Neighbors |
| LSTM | Long Short-Term Memory |
| MCI | mild cognitive impairment |
| MICT | Multimodal Input Cyclic Transformers |
| MLP | multilayer perceptron |
| ML | Machine Learning |
| MRI | Magnetic resonance imaging |
| MK-SVM | Multiple-Kernel SVM |
| PET | Positron emission tomography |
| qEEG | quantitative EEG |
| QDA | quadratic discriminant analysis |
| RBF | radial basis function |
| RF | Random Forests |
| RNNs | Recurrent Neural Networks |
| SNN | Self-Normalizing Neural Networks |
| SVM | Support vector machine |
| VAEs | Variational Autoencoders |

**References**

1. Feyissa AM, Tatum WO. Adult EEG. Handb Clin Neurol. 2019;160:103–24.

2. Benbadis SR, Beniczky S, Bertram E, MacIver S, Moshé SL. The role of EEG in patients with suspected epilepsy. Epileptic Disord. 2020 Apr 1;22(2):143–55.

3. de Gans CJ, Burger P, van den Ende ES, Hermanides J, Nanayakkara PWB, Gemke RJBJ, et al. Sleep assessment using EEG-based wearables - A systematic review. Sleep Med Rev. 2024 Aug;76:101951.

4. Janiesch C, Zschech P, Heinrich K. Machine learning and deep learning. Electron Markets. 2021 Sep 1;31(3):685–95.

5. LeCun Y, Bengio Y, Hinton G. Deep learning. Nature. 2015 May 28;521(7553):436–44.

6. Craik A, He Y, Contreras-Vidal JL. Deep learning for electroencephalogram (EEG) classification tasks: a review. J Neural Eng. 2019 Jun;16(3):031001.

7. Roy Y, Banville H, Albuquerque I, Gramfort A, Falk TH, Faubert J. Deep learning-based electroencephalography analysis: a systematic review. J Neural Eng. 2019 Aug 14;16(5):051001.

8. Rajwal S, Aggarwal S. Convolutional Neural Network-Based EEG Signal Analysis: A Systematic Review. Archives of Computational Methods in Engineering. 2023 Apr 10;30.

9. Li C, Qi Y, Ding X, Zhao J, Sang T, Lee M. A Deep Learning Method Approach for Sleep Stage Classification with EEG Spectrogram. International Journal of Environmental Research and Public Health. 2022 May 23;19(10):6322.

10. Graves A. Long Short-Term Memory. In: Graves A, editor. Supervised Sequence Labelling with Recurrent Neural Networks [Internet]. Berlin, Heidelberg: Springer; 2012 [cited 2024 Oct 16]. p. 37–45. Available from: https://doi.org/10.1007/978-3-642-24797-2_4

11. Penchina B, Sundaresan A, Cheong S, Martel A. Deep LSTM Recurrent Neural Network for Anxiety Classification from EEG in Adolescents with Autism. In 2020. p. 227–38.

12. Bethge D, Hallgarten P, Grosse-Puppendahl T, Kari M, Chuang L, Özdenizci O, et al. EEG2Vec: Learning affective EEG representations via variational autoencoders: 2022 IEEE International Conference on Systems, Man, and Cybernetics. 2022 IEEE International Conference on Systems, Man, and Cybernetics (SMC). 2022;3150–7.

13. Yao Y, Plested J, Gedeon T. Deep Feature Learning and Visualization for EEG Recording Using Autoencoders: 25th International Conference on Neural Information Processing, ICONIP 2018. Ozawa S, Leung ACS, Cheng L, editors. Neural Information Processing - 25th International Conference, ICONIP 2018, Proceedings. 2018;554–66.

14. Liu S, Cai W, Liu S, Zhang F, Fulham M, Feng D, et al. Multimodal neuroimaging computing: a review of the applications in neuropsychiatric disorders. Brain Informatics. 2015 Aug 29;2(3):167.

15. Rawnaque FS, Rahman KM, Anwar SF, Vaidyanathan R, Chau T, Sarker F, et al. Technological advancements and opportunities in Neuromarketing: a systematic review. Brain Inform. 2020 Sep 21;7(1):10.

16. Rabbani MHR, Islam SMR. Deep learning networks based decision fusion model of EEG and

fNIRS for classification of cognitive tasks. Cogn Neurodyn. 2024 Aug;18(4):1489–506.

17. Liao M, Duan H, Wang G. Application of Machine Learning Techniques to Detect the Children with Autism Spectrum Disorder. Journal of Healthcare Engineering. 2022;2022(1):9340027.

18. Rahman MA, Brown DJ, Mahmud M, Harris M, Shopland N, Heym N, et al. Enhancing biofeedback-driven self-guided virtual reality exposure therapy through arousal detection from multimodal data using machine learning. Brain Informatics. 2023 Jun 21;10(1):14.

19. Planke LJ, Gardi A, Sabatini R, Kistan T, Ezer N. Online Multimodal Inference of Mental Workload for Cognitive Human Machine Systems. Computers. 2021 Jun;10(6):81.

20. Farina FR, Emek-Savaş DD, Rueda-Delgado L, Boyle R, Kiiski H, Yener G, et al. A comparison of resting state EEG and structural MRI for classifying Alzheimer's disease and mild cognitive impairment. Neuroimage. 2020 Jul 15;215:116795.

21. Liu Y, Sivathamboo S, Goodin P, Bonnington P, Kwan P, Kuhlmann L, et al. Epileptic Seizure Detection Using Convolutional Neural Network: A Multi-Biosignal study. In: Proceedings of the Australasian Computer Science Week Multiconference 2020, ACSW 2020 [Internet]. Association for Computing Machinery (ACM); 2020 [cited 2024 Oct 16]. p. 37. Available from: https://research.monash.edu/en/publications/epileptic-seizure-detection-using-convolutional-neural-network-a-

22. Sirpal P, Kassab A, Pouliot P, Nguyen DK, Lesage F. fNIRS improves seizure detection in multimodal EEG-fNIRS recordings. J Biomed Opt. 2019 Feb;24(5):1–9.

23. Zhang J, Swinnen L, Chatzichristos C, Van Paesschen W, De Vos M. Learning Robust Representations of Tonic-Clonic Seizures With Cyclic Transformer. IEEE J Biomed Health Inform. 2024 Jun;28(6):3721–31.

24. Wadhera T. Multimodal Kernel-based discriminant correlation analysis data-fusion approach: an automated autism spectrum disorder diagnostic system. Phys Eng Sci Med. 2024 Mar 1;47(1):361–9.

25. Wu R, Li A, Xue C, Chai J, Qiang Y, Zhao J, et al. Screening for Mild Cognitive Impairment with Speech Interaction Based on Virtual Reality and Wearable Devices. Brain Sci. 2023 Aug 21;13(8):1222.

26. Amiri M, Raimondo F, Fisher PM, Hribljan MC, Sidaros A, Othman MH, et al. Multimodal Prediction of 3- and 12-Month Outcomes in ICU Patients with Acute Disorders of Consciousness. Neurocritical Care. 2023 Sep 11;40(2):718.

27. Ellis CA, Zhang R, Carbajal DA, Miller RL, Calhoun VD, Wang MD. Explainable Sleep Stage Classification with Multimodal Electrophysiology Time-series. Annu Int Conf IEEE Eng Med Biol Soc. 2021 Nov;2021:2363–6.

28. Abdollahpour M, Yousefi Rezaii T, Farzamnia A, Saad I. Transfer Learning Convolutional Neural Network for Sleep Stage Classification Using Two-Stage Data Fusion Framework. IEEE Access. 2020 Jan 1;8:180618–32.

29. Deligani RJ, Borgheai SB, McLinden J, Shahriari Y. Multimodal fusion of EEG-fNIRS: a mutual information-based hybrid classification framework. Biomed Opt Express. 2021 Mar 1;12(3):1635–50.

30. Jeong B, Lee J, Kim H, Gwak S, Kim YK, Yoo SY, et al. Multiple-Kernel Support Vector Machine for Predicting Internet Gaming Disorder Using Multimodal Fusion of PET, EEG, and Clinical Features. Front Neurosci [Internet]. 2022 Jun 30 [cited 2024 Oct 16];16. Available from: https://www.frontiersin.org/journals/neuroscience/articles/10.3389/fnins.2022.856510/full

31. Zhu J, Wang Y, La R, Zhan J, Niu J, Zeng S, et al. Multimodal Mild Depression Recognition Based on EEG-EM Synchronization Acquisition Network. IEEE Access. 2019;7:28196–210.

32. Chai J, Wu R, Li A, Xue C, Qiang Y, Zhao J, et al. Classification of mild cognitive impairment based on handwriting dynamics and qEEG. Comput Biol Med. 2023 Jan;152:106418.

33. Hosseini P, Tran T, Pompili D, Elisevich K, Soltanian-Zadeh H. Multimodal Data Analysis of Epileptic EEG and rs-fMRI via Deep Learning and Edge Computing. Artificial Intelligence in Medicine. 2020 Feb 1;104:101813.

34. Memarian N, Kim S, Dewar S, Engel J, Staba RJ. Multimodal data and machine learning for surgery outcome prediction in complicated cases of mesial temporal lobe epilepsy. Comput Biol Med. 2015 Sep;64:67–78.

35. Zhu J, Wang Y, La R, Zhan J, Niu J, Zeng S, et al. Multimodal Mild Depression Recognition Based on EEG-EMG Synchronization Acquisition Network. IEEE Access. 2019;7:28196–210.

36. Roy S, Kiral-Kornek I, Harrer S. ChronoNet: A Deep Recurrent Neural Network for Abnormal EEG Identification [Internet]. arXiv; 2018 [cited 2024 Nov 19]. Available from: http://arxiv.org/abs/1802.00308

37. Lee D, Kim B, Kim T, Joe I, Chong J, Min K, et al. A ResNet-LSTM hybrid model for predicting epileptic seizures using a pretrained model with supervised contrastive learning. Sci Rep. 2024 Jan 15;14(1):1319.